\documentclass[letterpaper,twocolumn,showpacs,prl,aps,superscriptaddress]{revtex4-1}
\usepackage[latin1]{inputenc}
\usepackage{bm}
\usepackage{amsmath}
\usepackage{graphicx}
\usepackage{amssymb}

\begin{document}

\title{Exact solution for a non-Markovian dissipative quantum dynamics}

\author{Luca Ferialdi}
\email{ferialdi@ts.infn.it}
\affiliation{Dipartimento di Fisica,
Universit\`a di Trieste, Strada Costiera 11, 34151 Trieste, Italy.
 \\ Istituto Nazionale di Fisica Nucleare,
Sezione di Trieste, Strada Costiera 11, 34151 Trieste, Italy.}

\author{Angelo Bassi}
\email{bassi@ts.infn.it}
\affiliation{Dipartimento di Fisica,
Universit\`a di Trieste, Strada Costiera 11, 34151 Trieste, Italy.
 \\ Istituto Nazionale di Fisica Nucleare,
Sezione di Trieste, Strada Costiera 11, 34151 Trieste, Italy.}

\begin{abstract}
We provide the exact analytic solution of the stochastic Schr\"odinger equation describing an harmonic oscillator interacting with a non-Markovian and dissipative environment. This result represents an arrival point in the study of non-Markovian dynamics via stochastic differential equations. It is also one of the few exactly solvable models, for infinite dimensional systems. We compute the Green's function; in the case of a free particle, and with an exponentially correlated noise, we discuss the evolution of Gaussian wave functions.
\end{abstract}


\maketitle

Non-Markovian quantum dynamics are rapidly becoming a popular field of research. From the theoretical point of view, there is a increasing interest in understanding the behavior of physical systems beyond the Markov approximation~\cite{Strunz:99,Adler:07,Breuer:08,meas}. This interest is triggered by novel experiments~\cite{exp,Liu:11}, which do not find explanation within the standard Markovian description. Moreover, people are already speaking of possible future applications, for example in describing energy transfer in biological systems~\cite{appl}.

There are different approaches to the description of (Markovian and non-Markovian) open quantum dynamics~\cite{hpz,Gindesperger:06,Piilo:08,noi}, among which stochastic differential equations (SDEs). In this approach, one modifies the Schr\"odinger equation by introducing new stochastic terms, which mimic the effect of an external agent (typically, the environment). This method has particular advantages: it allows for a description in terms of state vectors, instead of density matrices; it helps in finding exact analytic solutions for the dynamics, as in the case presented here; in some cases, it allows for faster numerical simulations. 

A particularly relevant example of SDE is the following~\cite{report,2021,222325,Chruscinski:92,Halliwell:95,Holevo:96,Bassi5:08,Bassi:08}:
\begin{equation}\label{eq:freepart}
\frac{d}{dt} \phi_t = \left[ - \frac{i}{\hbar} H + \sqrt{\lambda} q
w_t - \lambda q^2 \right] \phi_t\,;
\end{equation}
here, $q$ is the position operator of the particle, $H$ its quantum
Hamiltonian, $\lambda$ is a positive coupling constant, and $w_t$ is a Gaussian white noise. The equation does not preserve the norm of the state vector; this is not a problem, as at any time the vector can be appropriately normalized. This equation plays a key role in several fields. Within decoherence theory, it represents one of the possible unravelling of the Joos-Zeh master equation~\cite{jz}; within collapse models, it represents the simplest description of a quantum particle undergoing spontaneous collapses in space~\cite{report}; within the continued measurement theory, it describes a system whose position is continuously measured~\cite{meas}. In the following, we will focus our attention on collapse models, though our results apply also to the other fields.

If one tries to identify the noise $w_t$ with a physical field, one has to cope with the fact that it presents two unphysical features: it is an infinite-temperature field, due to the lack of dissipative terms, and has a white noise spectrum, i.e. each frequency contributes with the same weight. These features can be disturbing, and people worked in improving the model. The dissipative generalization of Eq.~(\ref{eq:freepart}) has been developed in~\cite{bassi3:05}; in the new dynamics, the noise is such that each system thermalizes to a fixed temperature, that of the field. The Markovian behavior has been generalized by studying the non-Markovian version of Eq.~(\ref{eq:freepart}), where the white noise is replaced by a general Gaussian (colored) noise~\cite{noi,Diosi:98,noi2}. 

In this Letter, we present a novel result~\cite{lungo}: the generalization of Eq.~(\ref{eq:freepart}) both to finite-temperature (dissipative) and colored (non-Markovian) noises, and its solution. The equation is the following:
\begin{eqnarray}\label{eq:gensde}
\frac{d}{dt}\phi_t&=&\left[-\frac{i}{\hbar}\left(H+\frac{\lambda\mu}{2}\{q,p\}\right)+\sqrt{\lambda}\left(q+i\frac{\mu}{\hbar}p\right)w_t\right.\nonumber\\
&&\left.-2\sqrt{\lambda}q\int_0^t ds\, D(t,s)\frac{\delta}{\delta w_s}\right]\phi_t\,.
\end{eqnarray}
The new terms depending on the momentum operator $p$ account for dissipation, whose strength is determined by the positive constant $\mu$. The temperature of the noise is determined by the relation: $T=\hbar^{2}(4mK_{\text{\tiny B}}\mu)^{-1}$.
The integral term of Eq.~\eqref{eq:gensde}, which involves the whole past history of the system, accounts for the non-Markovian behavior of the dynamics: $D(t,s)$ is the time correlation function of the noise, and $\frac{\delta}{\delta w_s}$ denotes a functional derivative with respect to $w_s$.

Eq.~(\ref{eq:gensde}) plays a central role, because it reproduces all previous models, under the appropriate limits. When $D(t,s)\rightarrow\delta(t-s)$, it reduces to the Markovian dissipative model analyzed in~\cite{bassi3:05}. When $\mu \rightarrow 0$, it reproduces the non-dissipative and non-Markovian model presented in~\cite{noi,noi2}; when both limits are taken, it reduces to Eq.~\eqref{eq:freepart}.

We succeeded in finding the general solution of Eq.~(\ref{eq:gensde}), for an harmonic oscillator~\cite{lungo}, which we will describe here below. Taking into account that only for equations containing terms which are most quadratic in $p$ and $q$, a general method for finding the solutions is known, our equation represents the most general equation that can be analytically solved (modulo physically unimportant terms). This matches with the fact that, from the physical point of view, the model represents an arrival point, since it involves both a non-Markovian and dissipative noise. Note that the model is similar to that of Hu, Paz, Zhang, describing non-Markovian quantum Brownian motion~\cite{hpz}.

{\it Solution.} The solution of Eq.~\eqref{eq:gensde} is given in terms of its Green's function $G(x,t;x_0,0)$. In~\cite{Diosi:97} it has been shown how $G(x,t;x_0,0)$ can be written via the path integral formalism as follows:
\begin{equation}\label{eq:propform}
G(x,t;x_0,0) \; = \; \int^{q(t)=x}_{q(0)=x_0} \mathcal{D}[q] \;
e^{\mathcal{S}[q]} \,,
\end{equation}
where $\mathcal{S}[q]$ is a non standard Action. For our model, it takes the expression:
\begin{widetext}
\begin{equation} \label{eq:action}
\mathcal{S}[q]  =  \int^t_0 ds \frac{i}{\hbar}\Bigg[ \frac{m}{2}\,
\dot{q}^2_s-m\lambda\mu\, q_s\,\dot{q}_s-\frac{m}{2}\Omega^2\,q^2_s+ m\sqrt{\lambda}\mu\, \dot{q}_s-A_s\,q_s+ q_s \int_0^s dr\, B(r,s)q_r-2m\lambda\mu\, q_s \int_0^s dr\, D(s,r)\dot{q}_r\Bigg]\,,
\end{equation}
with $\Omega^2=\omega^2-\lambda^2\mu^2$, where $\omega$ is the proper frequency of the oscillator, and
\begin{eqnarray}
\label{eq:A}A_s&=&i\hbar\sqrt{\lambda} w_s+m\lambda^{3/2}\mu^2 w_s+2m\lambda^{3/2}\mu^2\int_0^s\,dr D(r,s) w_r\,,\\
\label{eq:B}B(r,s)&=&\left(2m\lambda^2\mu^2+2i\hbar\lambda\right)D(r,s)+4m\lambda^2\mu^2\int_0^rdr' D(r,r')D(s,r')\,,
\end{eqnarray}
An important issue here arises. Since $\mathcal{S}[q]$ is  time-non-local, the standard Lagrange formalism cannot be applied; therefore we resorted to a time-non-local variational formalism  in~\cite{tnl}. The path integration in Eq.~\eqref{eq:action} was performed using the mid-point formulation of the Feynman polygonal approach~\cite{Feynman:65}. 
The calculation is long, and is duly reported in~\cite{lungo}; here we present only the final solution, which is an exact result. The Green's function reads:
\begin{equation}\label{eq:propexp}
G(x,t;x_0,0) = \sqrt{\frac{m}{2i\pi\hbar\, t\,
u(t)}}\exp\left[-\mathcal{A}_t
x_0^2-\tilde{\mathcal{A}}_tx^2+\mathcal{B}_tx_0x
+\mathcal{C}_tx_0+\mathcal{D}_tx +\mathcal{E}_t\right]\,,
\end{equation}
where $u(t)$ is a suitably defined function, unimportant for the subsequent analysis, $\tilde{\mathcal{A}}_t=-k\left(\dot{g}_t(t)-\lambda\mu\right)$, $k=\frac{im}{2\hbar}$, and
\begin{eqnarray}\label{eq:matha}
\mathcal{A}_t&=&k\left(\dot{f}_t(0)-\lambda\mu-2\lambda\mu\int_0^t ds\, D(0,s)f(s)\right)\,,\qquad
\mathcal{B}_t=k\left(\dot{f}_t(t)-\dot{g}_t(0)+2\lambda\mu\int_0^t ds\, D(0,s)g(s)\right)\,, \\
\mathcal{C}_t&=&-k\left(\dot{h}_t(0) +2\sqrt{\lambda}\mu\,w_0+\sqrt{\lambda}\mu\int_0^t\dot{w}_sf_t(s)ds+\int_0^t\frac{A_s}{m}f_t(s)ds-2\lambda\mu\int_0^t ds\, D(0,s)h(s)\right)\,,\\
\mathcal{D}_t&=&k\left(\dot{h}_t(t)+2\sqrt{\lambda}\mu\,w(t)-\sqrt{\lambda}\mu\int_0^t\dot{w}_sg_t(s)ds-\int_0^t\frac{A_s}{m}g_t(s)ds\right)\,,\\
\label{eq:mathe}\mathcal{E}_t&=&-k\left(\sqrt{\lambda}\mu\int_0^t\dot{w}_sh_t(s)ds+\int_0^t\frac{A_s}{m}h_t(s)ds-\lambda\mu^2\int_0^tw^2_s\,ds\right)\,,
\end{eqnarray}
where the differentiation is always done with respect to the variable between parenthesis.
Defining the following integro-differential operator:
\begin{eqnarray}
I[e(s)]&:=&\frac{m}{2}\, \ddot{e}(s)+\frac{m}{2}\left(\Omega^2+4\lambda\mu D(s,s)\right)\, e(s)-\frac{1}{2}\int_0^s dr\, B(r,s)e(r)\nonumber\\
&&-\frac{1}{2}\int_s^t dr\, B(s,r)e(r)-m\lambda\mu\int_0^s dr\, \frac{\partial D(r,s)}{\partial r}e(r)-m\lambda\mu\int_s^t dr\, \frac{\partial D(r,s)}{\partial s}e(r)\,,
\end{eqnarray}
\end{widetext}
one can prove that $f(s)$, $g(s)$, and $h(s)$ solve the following integro-differential equations:
\vspace{-0.0001cm}
\begin{eqnarray}
\label{eq:fans} I[f(s)] & = & m\lambda\mu\, D(0,s)\,\qquad
I[g(s)] = 0, \\
I[h(s)] & = & \frac{m\sqrt{\lambda}\mu}{2}\,\dot{w}(s)-\frac{A(s)}{2}\,, \label{eq:hans}
\end{eqnarray}
with suitable boundary conditions~\cite{lungo}. % $f(0)=1$, $f(t)=0$, $g(0)=0$, $g(t)=1$, $h(0)=0$, $h(t)=0$.

\vspace{0.17cm}
The Green's function~\eqref{eq:propexp} has some important features. The first one is that it has a Gaussian structure. This implies, first of all, that Gaussian states evolve into Gaussian states. It implies also that the density matrix has a Gaussian structure, because it is the average (an integral) with respect to a Gaussian noise, of states which evolve under a Gaussian-preserving dynamics. A second property is that the coefficients $\tilde{\mathcal{A}}$, $\mathcal{A}$ and $\mathcal{B}$ are deterministic, while only $\mathcal{C}$, $\mathcal{D}$, and $\mathcal{E}$ depend on the noise $w_t$ and thus evolve stochastically. The fact that the first three coefficients do not depend on the noise $w_t$ guarantees that the spread of a wave function evolves deterministically in time. 
From the point of view of collapse models, this is an important property, because it guarantees that any wave function collapses in space. Finally, one can check that, as expected, in the limit $\mu\rightarrow0$ the solution of the non-Markovian model is reproduced~\cite{noi}, while in the limit $D(t,s)\rightarrow\delta(t-s)$ one recovers the Markovian dissipative model~\cite{bassi3:05}. Considering both limits one obtains the Green's function of Eq.~\eqref{eq:freepart}. It is worthwhile stressing that the proof of all these properties relies on the fact that Eqs.~(\ref{eq:propexp})--(\ref{eq:hans}) represent the analytical solution of Eq.~(\ref{eq:gensde}).

The explicit expression of the Green's function depends on the solution of the three integro-differential equations~\eqref{eq:fans}-\eqref{eq:hans}. These equations can be exactly solved only for some correlation functions $D(t,s)$.  In order to give a quantitative example of the features of the model, we have solved Eqs.~\eqref{eq:fans}-\eqref{eq:hans} for the exponential correlation function and have studied the behavior of the spread of  Gaussian wave functions.

{\it Exponential correlation function. Wave function for a free particle.} 
Let us consider an exponential correlation function
\begin{equation}\label{eq:expcorr}
D(t,s)=(\gamma/2)e^{-\gamma |t-s|}\,
\end{equation}
which, besides allowing for an explicit solution of the equations, also represents a good physical example of a correlation function with a finite correlation time, given by $\gamma^{-1}$.
Using Eq.~(\ref{eq:expcorr}), one can show that the integro-differential equations~\eqref{eq:fans}-\eqref{eq:hans} can be transformed into fourth-order differential equations~\cite{Polyanin:08}. One can solve these equations for a generic quadratic potential (in particular, the harmonic oscillator~\cite{lungo}). For simplicity, here we consider a free particle, whose equation of $f(s)$ reads:
\begin{eqnarray}\label{eq:f4}
\ddddot{f}(s)-\left(\gamma^2+\lambda^2\mu^2-2\lambda\mu\gamma\right)\ddot{f}(s)&&\nonumber\\
+\left(4\lambda^2\mu^2\gamma^2+\frac{2i\hbar\lambda\gamma^2}{m} \right)f(s)&=&0\,.
\end{eqnarray}
The general solution is
\begin{equation}\label{eq:sol}
f_t(s) = \sum_{k=1}^{2} [ f_{t,k} \sinh \upsilon_k s + g_{t,k}
\cosh \upsilon_k s ]\,,
\end{equation}
 where the coefficients $f_{t,k}$, $g_{t,k}$ are determined by
the boundary conditions, and $\upsilon_1$, $\upsilon_2$ are the two roots
of the characteristic polynomial
associated to Eq.~\eqref{eq:f4}: 
\begin{eqnarray} \label{eq:gdsfsdasda}
\upsilon_{1,2}&=&\sqrt{\left((\gamma-\lambda\mu)^2\pm\zeta\right)/2}\,,\\
\zeta&=&\sqrt{(\gamma-\lambda\mu)^4-16\lambda^2\mu^2\gamma^2-\frac{8i\hbar\lambda\gamma^2}{m}}\,.
\end{eqnarray}
Two of the four boundary conditions are $f_t(0) = 1$ and $f_t(t)
= 1$, while the other two can be determined using a standard procedure explained in~\cite{noi2,Polyanin:08}:
\begin{eqnarray}
\ddot{f}(0)&=&\lambda^2\mu^2 f(0)+\lambda\mu\gamma\int_0^t dl\,
e^{-\gamma l}\dot{f}(l)\nonumber\\
&&+\left(\lambda^2\mu^2\gamma+\frac{i\hbar\lambda\gamma}{m}\right)\int_0^t dl\,
e^{-\gamma l}f(l)\,,\\
\ddot{f}(t)&=&\lambda^2\mu^2 f(t)-\lambda\mu\gamma\int_0^t dl\,
e^{-\gamma (t-l)}\dot{f}(l)\nonumber\\
&&+\left(\frac{3\lambda^2\mu^2\gamma}{2}+\frac{i\hbar\lambda\gamma}{m}\right)\int_0^t dl\,
e^{-\gamma (t-l)}f(l)\nonumber\\
&&-\frac{\lambda^2\mu^2\gamma}{2}\int_0^t dl\,
e^{-\gamma (t+l)}f(l)\,.
\end{eqnarray}
These four boundary conditions allow to determine the exact form of the coefficients $f_{t,k}$ and $g_{t,k}$; their expressions are too long to be written here, but can be easily worked out e.g. with Mathematica$^{\circledR}$. The same procedure here described can be applied also in order to find the analytic expressions for $g(s)$ and $h(s)$.  All the details of the calculation are reported in~\cite{lungo}.

Having found the explicit expression for every coefficient of the Green's function~(\ref{eq:propexp}), one can analyze the time evolution of wave functions. Particularly interesting, and easy to analyze, are Gaussian states, whose form---as previously discussed---is preserved by the dynamics. Accordingly, a wave function of the type:
\begin{equation}
\phi_t(x)=\exp[-\alpha_t x^2+\beta_tx+\gamma_t]\,,
\end{equation}
is solution of Eq.~(\ref{eq:gensde}), and the coefficients $\alpha_t$, $\beta_t$ and $\gamma_t$ evolve in time as follows:
\begin{eqnarray}
\alpha_t&=&\tilde{\mathcal{A}}_t-\frac{\mathcal{B}^2_t}{4(\alpha_0+\mathcal{A}_t)},
\quad
\beta_t = -\frac{\mathcal{C}_t+\beta_0}{4(\alpha_0+\mathcal{A}_t)}+\mathcal{D}_t \nonumber \\
\gamma_t&=&\gamma_0+\mathcal{E}_t+\frac{(\mathcal{C}_t+\beta_0)^2}{4(\alpha_0+\mathcal{A}_t)}\,.
\end{eqnarray}
As anticipated, one can easily see that the evolution of the spread both in position and in momentum is deterministic since the parameters $\tilde{\mathcal{A}}$, $\mathcal{A}$ and $\mathcal{B}$ do not depend on the noise.
In particular, we focus our attention on the behavior of the spread in position: $\sigma(t) =
1/2\sqrt{\alpha^{\text{\tiny R}}_t}$ (the apex $\text{\tiny R}$ denotes the real part) in the case of the exponential
correlation function~\eqref{eq:expcorr}.
\begin{figure}[]
\begin{center}
\includegraphics[width=8.5cm]{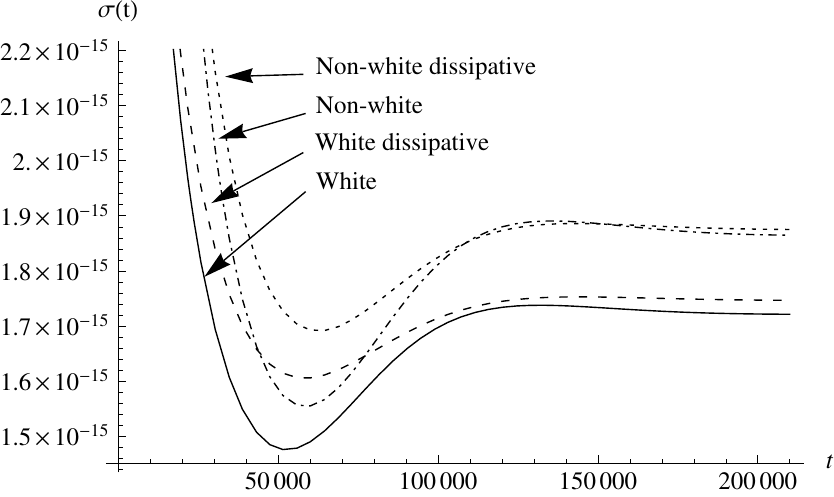}
\caption{Time evolution of the spread in position $\sigma(t)$ of a Gaussian wave function, in the case of an exponential correlation function for the noise. The initial spread is $\sigma(0)=1$ m. The other parameters have been chosen as follows: $m=1$ kg, $\lambda=3\times10^{24}$ m$^{-2}$ s$^{-1}$, $\mu=10^{-30}$ m$^2$ (dissipative models), $\gamma=10^{-4}$ s$^{-1}$ (non-white models). Time is measured in seconds, distances in meters. For very small values of $\sigma(0)$ the wave function spreads out instead of shrinking, reaching the asymptotic value displayed in the figure (see also~\cite{raviola}).} \label{fig:1}
\end{center}
\end{figure}

Fig.~\ref{fig:1} shows the comparison of the evolution of the spread according to the four versions of the model (white~\cite{Bassi2:05}, white dissipative~\cite{bassi3:05}, non-white~\cite{noi}, non-white dissipative). First of all one can see that the behavior is qualitatively the same for every model: the spread decreases in time, reaching an asymptotic finite value. From the quantitative point of view one can see that both in the non-Markovian (dot-dashed line) and in the dissipative (dashed line) models the shrinking of the wave function is slower than in the white noise case (solid line). As a consequence, when these two effects are combined (dotted line), the spread decreases even more slowly. This is easy to understand. In the Markovian case, all frequencies contribute to the collapse of the wave function, while in the non-Markovian case the high-frequency components are suppressed. This slows the collapse process. Secondly, a finite temperature noise is less energetic than an infinite temperature noise, therefore in the dissipative case the collapse is slower than in the non-dissipative case.

From the point of view of collapse models, an interesting physical question is whether a field with `typical' cosmological values for the temperature ($T \simeq 2.73 \text{K}$, that of the CMBR) and for the spectrum (cutoff at $\gamma \sim 10^{10}-10^{11} \text{Hz}$, that of the CMBR, the relic neutrino background, and the relic gravitational background), can collapse the wave function efficiently. This question has been answered positively in~\cite{noi3}: a noise with typical cosmological features can collapse the wave function fast enough to guarantee the emergence of classical properties at the macroscopic level.

{\it Acknowledgements.} The authors acknowledge partial financial support from MIUR (PRIN 2008), INFN, COST (MP1006) and the John Templeton Foundation project \lq Quantum Physics and the Nature of Reality\rq.

\def\polhk#1{\setbox0=\hbox{#1}{\ooalign{\hidewidth
  \lower1.5ex\hbox{`}\hidewidth\crcr\unhbox0}}} \def\cprime{$'$}
  \def\cprime{$'$}

\end{document}